\documentclass[aps,prd,onecolumn,showpacs,showkeys,superscriptaddress,preprintnumbers,floatfix,reprint]{revtex4}
\usepackage{graphicx}
\usepackage{amsmath}
\usepackage{bm}
\usepackage{yhmath}
\usepackage{hyperref}
\usepackage{subfigure}
\usepackage{color}
\usepackage{cases}
\usepackage{subfigure}
\usepackage{dcolumn,booktabs,bm}

\usepackage{amsfonts,amssymb,stmaryrd,latexsym,amsmath}
\usepackage{textcomp}

\usepackage{array}
\newcounter{RomanNumber}

\newcommand{\lyxmathsym}[1]{\ifmmode\begingroup\def\b@ld{bold}
  \text{\ifx\math@version\b@ld\bfseries\fi#1}\endgroup\else#1\fi}

\def\mud{{\lambda}}

\allowdisplaybreaks

\begin{document}

\title{Magnetic moments of the doubly charmed and bottomed baryons}

\author{Hao-Song Li}\email{haosongli@pku.edu.cn}\affiliation{School of Physics and State Key Laboratory of Nuclear Physics and Technology, Peking University, Beijing 100871, China}

\author{Lu Meng}\email{lmeng@pku.edu.cn}
\affiliation{School of Physics, Peking University, Beijing 100871,
China}
\author{Zhan-Wei Liu}\email{liuzhanwei@lzu.edu.cn}
\affiliation{School of Physical Science and Technology, Lanzhou
University, Lanzhou 730000, China}

\author{Shi-Lin Zhu}\email{zhusl@pku.edu.cn}\affiliation{School of Physics and State Key Laboratory of Nuclear Physics and Technology, Peking University, Beijing 100871, China}\affiliation{Collaborative Innovation Center of Quantum Matter, Beijing 100871, China}

\begin{abstract}

The chiral corrections to the magnetic moments of the
spin-$\frac{1}{2}$ doubly charmed baryons are systematically
investigated up to next-to-next-to-leading order with heavy baryon
chiral perturbation theory (HBChPT). The numerical results are
calculated up to next-to-leading order:
$\mu_{\Xi^{++}_{cc}}=-0.25\mu_{N}$,
$\mu_{\Xi^{+}_{cc}}=0.85\mu_{N}$,
$\mu_{\Omega^{+}_{cc}}=0.78\mu_{N}$. We also calculate the magnetic
moments of the other doubly heavy baryons, including the doubly
bottomed baryons (bbq) and the doubly heavy baryons containing a
light quark, a charm quark and a bottom quark ($\{bc\}q$ and
$[bc]q$): $\mu_{\Xi^{0}_{bb}}=-0.84\mu_{N}$,
$\mu_{\Xi^{-}_{bb}}=0.26\mu_{N}$,
$\mu_{\Omega^{-}_{bb}}=0.19\mu_{N}$,
$\mu_{\Xi^{+}_{\{bc\}q}}=-0.54\mu_{N}$,
$\mu_{\Xi^{0}_{\{bc\}q}}=0.56\mu_{N}$,
$\mu_{\Omega^{0}_{\{bc\}q}}=0.49\mu_{N}$,
$\mu_{\Xi^{+}_{[bc]q}}=0.69\mu_{N}$,
$\mu_{\Xi^{0}_{[bc]q}}=-0.59\mu_{N}$,
$\mu_{\Omega^{0}_{[bc]q}}=0.24\mu_{N}$.

\end{abstract}

\maketitle

\thispagestyle{empty}

\section{Introduction}\label{Sec1}

SELEX Collaboration first reported evidence for the doubly charmed
baryon $\Xi^{+}_{cc}$(3520) in the decay mode
$\Xi^{+}_{cc}\rightarrow\Lambda_{c}^{+}K^{-}\pi^{+}$ with the mass
$M_{\Xi^{+}_{cc}}=3519\pm1\rm{MeV}$~\cite{Mattson:2002vu}, although
other experimental collaborations like FOCUS~\cite{Ratti:2003ez},
BABAR~\cite{Aubert:2006qw} and Belle~\cite{Chistov:2006zj} did not
find any evidence of the doubly charmed baryons. Recently, LHCb
collaboration observed $\Xi^{++}_{cc}$ in the
$\Lambda_{c}^{+}K^{-}\pi^{+}\pi^{+}$ mass spectrum with the mass
$M_{\Xi^{++}_{cc}}=3621.40\pm0.72 (\rm stat)\pm0.27 (\rm
syst)\pm0.14 (\Lambda^{+}_{c}) \rm{MeV}$~\cite{LHCb}.

In the past decade, there have been many investigations of the
doubly charmed baryon masses
\cite{Bagan:1992za,Roncaglia:1995az,SilvestreBrac:1996bg,Ebert:1996ec,Tong:1999qs,Itoh:2000um,Gershtein:2000nx,
Kiselev:2001fw,Kiselev:2002iy,Narodetskii:2001bq,Lewis:2001iz,Faessler:2001mr,Ebert:2002ig,Mathur:2002ce,Flynn:2003vz,Vijande:2004at,Chiu:2005zc,
Migura:2006ep,Albertus:2006ya,Liu:2007fg,Roberts:2007ni,Valcarce:2008dr,Liu:2009jc,Namekawa:2012mp,Alexandrou:2012xk,
Aliev:2012ru,Aliev:2012iv,Namekawa:2013vu,Sun:2014aya,Chen:2015kpa,Sun:2016wzh,
Shah:2016vmd,Chen:2016spr,Kiselev:2017eic,Chen:2017sbg}. However,
the electromagnetic form factors, especially the magnetic moments
play a pivotal role in describing the inner structures of hadrons.
In the quark-model, the doubly charmed baryons are just like the
light baryons with two light quarks replaced by two charm quarks.
The magnetic moments of doubly charmed baryons were first
investigated by Lichtenberg in Ref.~\cite{Lichtenberg:1976fi} with
nonrelativistic qurak model. Since then, more elaborate quark models
have been developed to study the magnetic moments of doubly charmed
baryons. In Ref.~\cite{SilvestreBrac:1996bg}, various static
properties, including magnetic moments were studied within
non-relativistic quark model using the Faddeev formalism.
magnetic moments were also evaluated in the relativistic quark
model~\cite{JuliaDiaz:2004vh,Faessler:2006ft}. In Ref.~\cite{Branz:2010pq}, the radiative decays of double heavy baryons were studied in a relativistic constituent three-quark model including hyperfine mixing.

Besides the quark models, the magnetic moments of the doubly charmed
baryons have been studied with other approaches, such as the MIT bag
model \cite{Bose:1980vy,Bernotas:2012nz}, the Dirac equation
formalism \cite{Jena:1986xs}, the Skyrmion model \cite{Oh:1991ws},
the hyper central description of the three-body system
\cite{Patel:2008xs} and lattice QCD \cite{Can:2013zpa,Can:2013tna}.
In Refs.~\cite{Can:2013zpa,Can:2013tna}, the authors studied the
electromagnetic properties of  baryons in 2+1 flavor lattice QCD.
They found that the magnetic moments of the singly charmed baryons
are dominantly determined by the light quarks, while the charm
quarks play a more important role in the doubly charmed baryons,
which is confirmed in this paper.

Unfortunately most of the above models miss the chiral corrections.
The Goldstone boson cloud effect can be taken into account through
Chiral perturbation theory (ChPT)~\cite{Weinberg:1978kz}, which
organizes the low-energy interactions order by order. Since the
baryon mass $M$ does not vanish in the chiral limit, the convergence
of the chiral expansion is destroyed by the large energy scale $M$.
To overcome the above difficulty, heavy baryon chiral perturbation
theory (HBChPT) was
proposed~\cite{Jenkins:1990jv,Jenkins:1992pi,Bernard:1992qa,Bernard:1995dp},
which has been successfully used in the investigation of baryons.
For the doubly charmed baryons, the two charmed quarks are so heavy
that they can be treated as spectators. Thus, the remaining light
quark dominates the chiral dynamics of the doubly charmed baryons.

In this work, we will investigate the magnetic moments of the
spin-$\frac{1}{2}$ doubly charmed or bottomed baryons with HBChPT.
Right now, there does not exist any experimental measurement of the
magnetic moments of the doubly charmed baryons. We use quark model
to estimate the corresponding low energy constants (LECs) and
calculate the chiral corrections to the magnetic moments order by
order.  The numerical results are presented up to next-to-leading
order while the analytical results are calculated to
next-to-next-to-leading order.

Our work is organized as follows. In Section \ref{Sec3}, we discuss
the electromagnetic form factors of the spin-$\frac{1}{2}$ doubly
charmed baryons. In Section \ref{Sec2}, we introduce the effective
chiral Lagrangians. We calculate the chiral corrections to the
magnetic moments order by order in Section \ref{secFormalism} and
present our numerical results in Section \ref{Sec6}. A short summary
is given in Section \ref{Sec7}. We collect
the coefficients of the loop corrections in the
Appendix~\ref{appendix-B}.

\section{Electromagnetic form factors of spin-$\frac{1}{2}$ doubly charmed baryon baryon }\label{Sec3}

For the spin-$\frac{1}{2}$ doubly charmed baryons, the matrix
elements of the electromagnetic current is similar to that of the
nucleon,
\begin{equation}
<\Psi(p^{\prime})|J_{\mu}|\Psi(p)>=e\bar{u}(p^{\prime})\mathcal{O}_{\mu}(p^{\prime},p)u(p),
\end{equation}
with
\begin{equation}
\mathcal{O}_{\mu}(p^{\prime},p)=\frac{1}{M_H}[P_{\mu}G_{E}(q^{2})+\frac{i\sigma_{\mu\nu}q^{\nu}}{2}G_{M}(q^{2})].
\label{eq_new_current}
\end{equation}
where $P=\frac{1}{2}(p^{\prime}+p)$, $q=p^{\prime}-p$, $M_{H}$ is
the doubly charmed baryon mass.

As the doubly charmed baryons are very heavy compared to the chiral
symmetry breaking scale, we adopt the heavy-baryon formulation. In
the heavy baryon limit, the spin-$\frac{1}{2}$ doubly charmed baryon
field $B$ can be decomposed into the large component $H$ and the
small component $L$.
\begin{equation}
B=e^{-iM_{H}v\cdot x}(H+L),
\end{equation}
\begin{equation}
H=e^{iM_{H}v\cdot x}\frac{1+v\hspace{-0.5em}/}{2}B,~
L=e^{iM_{H}v\cdot x}\frac{1-v\hspace{-0.5em}/}{2}B,
\end{equation}
where $v_{\mu}=(1,\vec{0})$ is the velocity of the baryon. Now the
doubly charmed baryon matrix elements of the electromagnetic current
$J_{\mu}$ read
\begin{equation}
<H(p^{\prime})|J_{\mu}|H(p)>=e\bar{u}(p^{\prime})\mathcal{O}_{\mu}(p^{\prime},p)u(p)\label{eq:ocurrent}.
\end{equation}
The tensor $\mathcal{O}_{\mu}$ can be parameterized in terms of
electric and magnetic form factors.
\begin{eqnarray}
\mathcal{O}_{\mu}(p^{\prime},p)=v_{\mu}G_{E}(q^{2})+\frac{[S^{\mu},S^{\nu}]q^{\nu}}{M_H}G_{M}(q^{2}),
\label{eq_newnew_current}
\end{eqnarray}
where $G_{E}(q^{2})$ is the electric form factor and $G_{M}(q^{2})$
is the magnetic form factor. When $q^2=0$, we obtain the charge
($Q$) and magnetic moment ($\mu_{H}$),
\begin{eqnarray}
Q=G_{E}(0), \mu_{H}=\frac{e}{2M_H}G_{M}(0).
\label{eq_magneticcurrent}
\end{eqnarray}
\section{Chiral Lagrangians}\label{Sec2}

\subsection{The strong interaction chiral Lagrangians}

To calculate the chiral corrections to the magnetic moment, we
construct the relevant chiral Lagrangians. We follow
Refs.~\cite{Li:2016ezv,Scherer:2002tk,Bernard:1995dp} to define the
basic chiral effective Lagrangians of the pseudoscalar mesons.

The spin-$\frac{1}{2}$ doubly charmed baryon field reads
\begin{equation}
\Psi=\left(\begin{array}{c}
\Xi_{cc}^{++}\\
\Xi_{cc}^{+}\\
\Omega_{cc}^{+}
\end{array}\right)\Rightarrow\left(\begin{array}{c}
ccu\\
ccd\\
ccs
\end{array}\right).
\end{equation}
The leading order pseudoscalar meson and doubly charmed baryon
interaction Lagrangians read
\begin{eqnarray}
\mathcal{L}^{(1)}=\bar{\Psi}(iD\hspace{-0.6em}/-M_H)\Psi,
\label{Eq:baryon01}\\
\mathcal{L}_{{\rm
int}}^{(1)}=\frac{\tilde{g}_{A}}{2}\bar{\Psi}\gamma^{\mu}\gamma_{5}u_{\mu}\Psi
,\label{Eq:baryon02}
\end{eqnarray}
where $M_H$ is doublely charmed baryon mass,
\begin{eqnarray}
D_{\mu}\Psi&=&\partial_{\mu}\Psi+[\Gamma_{\mu},\Psi].
\end{eqnarray}
We also need the second order pseudoscalar meson and doubly charmed
baryon interaction Lagrangians. Recall that for SU(3) group
representations,
\begin{eqnarray}
3\otimes\bar{3} & = & 1\oplus8\label{Eq:flavor1},\\
8\otimes8 & = &
1\oplus8_{1}\oplus8_{2}\oplus10\oplus\bar{10}\oplus27.\label{Eq:flavor2}
\end{eqnarray}
Both $u_{\mu}$ and $u_{\nu}$ transform as the adjoint
representation. When the product of $u_{\mu}$ and $u_{\nu}$ belongs
to the $8_1$ and $8_2$ flavor representations, we can write down two
independent interaction terms of the second order pseudoscalar meson
and baryon Lagrangians:
\begin{eqnarray}
\hat{\mathcal{L}}_{\rm int}^{(2)}&=&\frac{ig_{h1}}{4M_{B}}
\bar{\Psi}\sigma^{\mu\nu}[u_{\mu},
u_{\nu}]\Psi+\frac{ig_{h2}}{4M_{B}}
\bar{\Psi}\sigma^{\mu\nu}\{u_{\mu}, u_{\nu}\}\Psi ,
\label{Eq:baryon03}
\end{eqnarray}
where the superscript denotes the chiral order, $M_B$ is the nucleon
mass and $g_{h1,h2}$ are the coupling constants. The $g_{h2}$ term
vanishes because of anti-symmetric lorentz structure. Thus, there is
only one linearly independent low energy constant (LEC) $g_{h1}$
which contributes to the present investigations of the doubly
charmed baryon magnetic moments up to $\mathcal{O}(p^4)$.

In the framework of HBChPT, the leading order nonrelativistic
pseudoscalar meson and doubly charmed baryon Lagrangians read
\begin{equation}
\mathcal{L}_{0}^{(1)}=\bar{H}(iv\cdot D)H, \label{Eq:baryon1}
\end{equation}
\begin{equation}
\mathcal{L}_{\rm int}^{(1)}=\tilde{g}_{A}{\rm Tr}\bar{H}S^{\mu}u_{\mu}H,
\label{Eq:baryon2}
\end{equation}
where $\mathcal{L}_{0}^{(1)}$ and $\mathcal{L}_{\rm int}^{(1)}$ are
the free and interaction parts respectively. $S_{\mu}$ is the
covariant spin-operator. We do not consider the mass differences
among different doubly charmed baryons. We estimated the $\phi H H$
coupling $\tilde{g}_{A}=0.5$ with the help of quark model in Section
\ref{Sec6}. For the pseudoscalar meson masses, we use
$m_{\pi}=0.140$ GeV, $m_{K}=0.494$ GeV, and $m_{\eta}=0.550$ GeV. We
use the nucleon masses $M_B=0.938\rm{GeV}$.

The second order pseudoscalar meson and baryon nonrelativistic
Lagrangians read
\begin{eqnarray}
\hat{\mathcal{L}}_{\rm int}^{(2)}&=&\frac{g_{h1}}{2M_{B}}
\bar{H}[S^\mu,S^\nu][u_{\mu}, u_{\nu}]H
.\label{Eq:HHUU}
\end{eqnarray}
The above Lagrangians contribute to the doubly charmed baryon
magnetic moments in diagram (e) of Fig.~\ref{fig:allloop}.

\subsection{The electromagnetic chiral Lagrangians at $\mathcal{O}(p^{2})$}

The lowest order $\mathcal{O}(p^{2})$ Lagrangian contributes to the
magnetic moments of the doubly charmed baryons at the tree level
\begin{equation}
\mathcal{L}_{\mu_{H}}^{(2)}=a_{1}\frac{-i}{4M_{B}}\bar{H}[S^{\mu},S^{\nu}]\hat{F}_{\mu\nu}^{+}H+a_{2}\frac{-i}{4M_{B}}\bar{H}[S^{\mu},S^{\nu}]H{\rm
Tr}(F_{\mu\nu}^{+}), \label{Eq:MM1}
\end{equation}
where the coefficients $a_{1,2}$ are the LECs. The chirally
covariant QED field strength tensor $F_{\mu\nu}^{\pm}$ is defined as
\begin{eqnarray} \nonumber
F_{\mu\nu}^{\pm} & = & u^{\dagger}F_{\mu\nu}^{R}u\pm
uF_{\mu\nu}^{L}u^{\dagger},\\
F_{\mu\nu}^{R} & = &
\partial_{\mu}r_{\nu}-\partial_{\nu}r_{\mu}-i[r_{\mu},r_{\nu}],\\
F_{\mu\nu}^{L} & = &
\partial_{\mu}l_{\nu}-\partial_{\nu}l_{\mu}-i[l_{\mu},l_{\nu}],
\end{eqnarray}
where $r_{\mu}=l_{\mu}=-eQ_HA_{\mu}$ and $Q_H=\rm{diag}(2,1,1)$. The
operator $\hat{F}_{\mu\nu}^{+}=F_{\mu\nu}^{+}-\frac{1}{3}\rm
Tr(F_{\mu\nu}^{+})$ is traceless and transforms as the adjoint
representation. Recall that the direct product $3\otimes\bar{3} =
1\oplus8$ . Therefore, there are two independent interaction terms
in the $\mathcal{O}(p^{2})$ Lagrangians for the magnetic moments of
the doubly charmed baryons.

\subsection{The higher order electromagnetic chiral Lagrangians }

To calculate the magnetic moments to $\mathcal{O}(p^{3})$, we also
need the $\mathcal{O}(p^{4})$ electromagnetic chiral Lagrangians at
the tree level. Recalling flavor representation in
Eqs.~(\ref{Eq:flavor1}), (\ref{Eq:flavor2}) and considering that we
only need the leading-order terms of the fields $F_{\mu\nu}^{+}$ and
$\chi^{+}$ which are diagonal matrices, only three independent terms
contribute to the magnetic moments of the doubly charmed baryons up
to $\mathcal{O}(p^{3})$,
\begin{eqnarray}
\mathcal{L}_{\mu_{H}}^{(4)}&=&d_{1}\frac{-i}{4M_{B}}\bar{H}[S^{\mu},S^{\nu}]H{\rm
Tr}(\chi^{+}F_{\mu\nu}^{+})+d_{2}\frac{-i}{4M_{B}}\bar{H}[S^{\mu},S^{\nu}]\{F_{\mu\nu}^{+},\chi^{+}\}H\nonumber
\\&&+d_{3}\frac{-i}{4M_{B}}\bar{H}[S^{\mu},S^{\nu}]\chi^{+}H{\rm Tr}(F_{\mu\nu}^{+})\label{Eq:MM3}
\end{eqnarray}
where $\chi^{+}$=diag(0,0,1) at the leading order and the factor
$m_{s}$ has been absorbed in the LECs $d_{1,2,3}$.

There are two more terms which also contribute to the doubly charmed
baryon magnetic moments.
\begin{eqnarray}
\mathcal{L^{\prime}}_{\mu_{H}}^{(4)}&=&a_{1}^{\prime}\frac{-i}{4M_{B}}\bar{H}[S^{\mu},S^{\nu}]F_{\mu\nu}^{+}H{\rm
Tr}(\chi^{+})+a_{2}^{\prime}\frac{-i}{4M_{B}}\bar{H}[S^{\mu},S^{\nu}]H{\rm
Tr}(F_{\mu\nu}^{+}){\rm Tr}(\chi^{+})
\end{eqnarray}
However, their contributions can be absorbed through the
renomalization of the LECs $a_{1,2}$, i.e.
\begin{eqnarray}
a_{1}&\rightarrow&a_{1}+{\rm Tr}(\chi^{+})a_{1}^{\prime},\\
a_{2}&\rightarrow&a_{2}+{\rm Tr}(\chi^{+})a_{2}^{\prime}.
\end{eqnarray}

\section{Formalism up to one-loop level}\label{secFormalism}

We follow the standard power counting scheme as in Ref.
\cite{Li:2017vmq}. The chiral order $D_{\chi}$ is given
by~\cite{Ecker:1994gg}
\begin{equation}
D_{\chi}=4N_{L}-2I_{M}-I_{B}+\sum_{n}nN_{n}, \label{Eq:Power
counting}
\end{equation}
where $N_{L}$ is the number of loops, $I_{M}$ is the number of the
internal pion lines, $I_{B}$ is the number of the internal baryon
lines and $N_{n}$ is the number of the vertices from the $n$th order
Lagrangians. The chiral order of the magnetic moments $\mu_{H}$ is
$(D_\chi-1)$ based on Eq. (\ref{eq_magneticcurrent}).

\begin{figure}
\centering
\includegraphics[width=0.6\hsize]{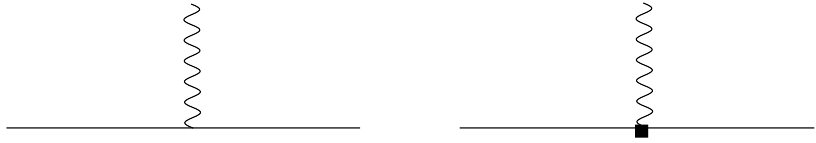}
\caption{The $\mathcal{O}(p^{2})$ and $\mathcal{O}(p^{4})$ tree
level diagrams where the doubly charmed baryon is denoted by the
 solid line. The left dot and the right black square
represent second- and fourth-order couplings respectively.}
\label{fig:tree}
\end{figure}

We assume the exact isospin symmetry with $m_{u}=m_{d}$ throughout
this work. The tree-level Lagrangians in Eqs.
~(\ref{Eq:MM1}),(\ref{Eq:MM3}) contribute to the doubly charmed
baryon magnetic moments at $\mathcal{O}(p^{1})$ and
$\mathcal{O}(p^{3})$ as shown in Fig.~\ref{fig:tree}. The
Clebsch-Gordan coefficients for the various doubly charmed baryons
are collected in Table~\ref{Magnetic moments}. All doubly charmed
baryon magnetic moments are given in terms of $a_{1}$, $a_{2}$,
$d_{1}$, $d_{2}$ and $d_{3}$.

\begin{figure}[tbh]
\centering
\includegraphics[width=0.9\hsize]{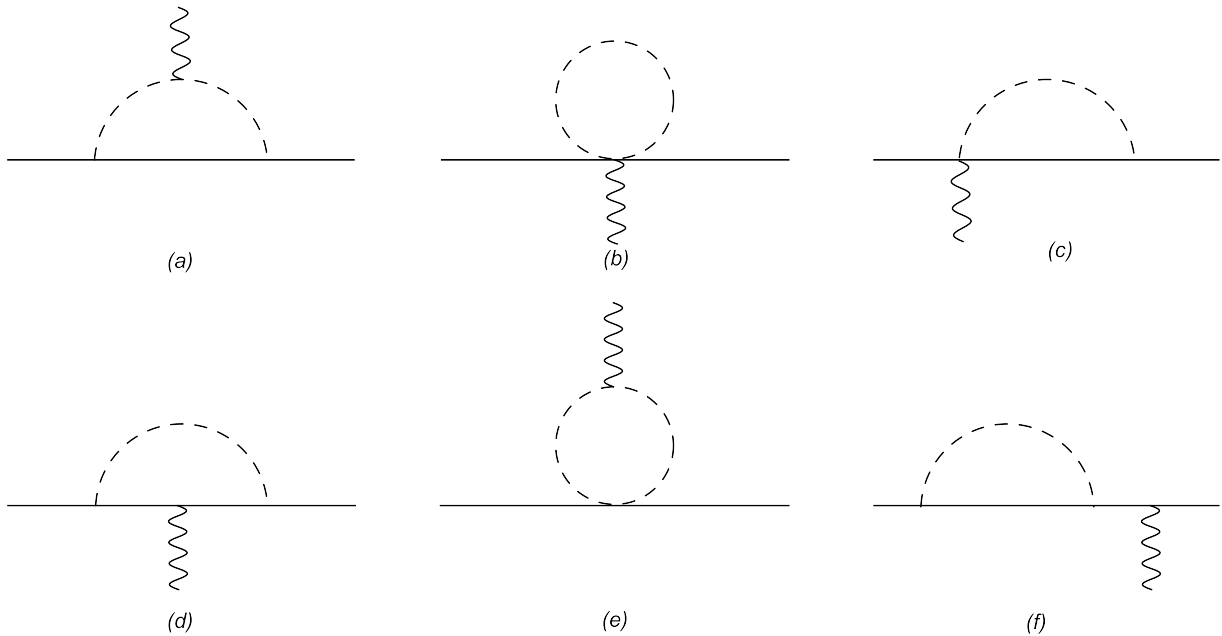}
\caption{The one-loop diagrams where the doubly charmed baryon is
denoted by the solid line. The dashed and wiggly lines represent the
pseudoscalar meson and photon respectively.}\label{fig:allloop}

\end{figure}

There are six Feynman diagrams contribute to the doubly charmed
baryon magnetic moments at one-loop level as shown in
Fig.~\ref{fig:allloop}. All the vertices in these diagrams come from
Eqs.~(\ref{Eq:baryon2}-\ref{Eq:MM1}). In diagrams (a), the meson
vertex is from the strong interaction terms while the photon vertex
is from the meson photon interaction term. In diagram (b), the
photon-meson-baryon vertex is from the $\mathcal{O}(p^{2})$ tree
level magnetic moment interaction in Eq.~(\ref{Eq:MM1}). In diagram
(c), the two vertices are from the strong interaction and seagull
terms respectively. In diagrams (d), the meson vertex is from the
strong interaction terms in Eq.~(\ref{Eq:baryon2}) while the photon
vertex from the $\mathcal{O}(p^{2})$ tree level magnetic moment
interaction in Eq.~(\ref{Eq:MM1}). In diagram (e), the meson-baryon
vertex is from the second order pseudoscalar meson and baryon
Lagrangian in Eq.~(\ref{Eq:HHUU}) while the photon vertex is also
from the meson photon interaction term. In diagram (f), the meson
vertex is from the strong interaction terms while the photon vertex
is from the $\mathcal{O}(p^{2})$ tree level magnetic moment
interaction in Eq.~(\ref{Eq:MM1}).

The diagram (a) contributes to the tensor $e \mathcal O_{\mu}$ in
Eq.~(\ref{eq:ocurrent}) at $\mathcal{O}(p^{3})$ while the diagrams
(b-f) contribute at $\mathcal{O}(p^{4})$. The diagram (c) vanishes
in the heavy baryon mass limit. In particular,
\begin{eqnarray}
J_{c}\propto\int\frac{d^{d}l}{(2\pi)^{d}}\frac{i}{l^{2}-m^{2}+i\epsilon}(S\cdot
l)\frac{i}{v\cdot l+i\epsilon}S^{\mu}\propto&S\cdot v=0.
 \end{eqnarray}
In other words, diagram (c) does not contribute to the magnetic
moments in the leading order of the heavy baryon expansion. The
diagram (f) indicates the corrections from the wave function
renormalization.

Summing all the contributions to the  doubly charmed baryon magnetic
moments in Fig.~\ref{fig:allloop}, the leading and next-to-leading
order loop corrections  can be expressed as
\begin{eqnarray}
\mu_{H}^{(2,\rm loop)}& = &-\sum_{\phi=\pi,K}\frac{\tilde{g}_{A}^2m_{\phi}M_{N}\beta_{a}^{\phi}}{64\pi f_{\phi}^2},\label{eq:mu2Loop}\\
\mu_{H}^{(3,\rm loop)}& =
&\sum_{\phi=\pi,K}[\frac{\beta^\phi_bm_{\phi}^2\ln
\frac{m_{\phi}^2}{\lambda ^2}}{128\pi^2 f_{\phi}^2}+
\frac{\beta^\phi_e m_{\phi}^2\ln \frac{m_{\phi}^2}{\lambda
^2}}{16\pi^2 f_{\phi}^2}]+\sum_{\phi=\pi,K,\eta}[\frac{-\beta^\phi_d
\tilde{g}_{A}^2m_{\phi}^2}{512\pi^2 f_{\phi}^2}(\ln
\frac{m_{\phi}^2}{\lambda ^2}-2)+ \frac{-3\beta^\phi_f
\tilde{g}_{A}^2m_{\phi}^2\ln \frac{m_{\phi}^2}{\lambda ^2}}{256\pi^2
f_{\phi}^2}] \label{eq:mu3Loop}
\end{eqnarray}
where $\mud=4\pi f_{\pi}$ is the renormalization scale. Here, we use
the number $n$ within the parenthesis in the superscript of $X^{(n,
...)}$ to indicate the chiral order of $X$. $\beta^\phi_{a-f}$ arise
from the corresponding diagrams in Fig.~\ref{fig:allloop}. We
collect their explicit expressions in Tables~\ref{table:abd} and
\ref{table:ef} in the Appendix \ref{appendix-B}.

With the low energy counter terms and loop contributions
(\ref{eq:mu2Loop}, \ref{eq:mu3Loop}), we obtain the magnetic
moments,
\begin{equation}
\mu_{H}=\left\{\mu_{H}^{(1)}\right\}+\left\{\mu_{H}^{(2,\rm
loop)}\right\}+\left\{\mu_{H}^{(3,\rm tree)}+\mu_{H}^{(3,\rm
loop)}\right\}
\end{equation}
where $\mu_{H}^{(1)}$ and $\mu_{H}^{(3,\rm tree)}$ are the
tree-level magnetic moments from Eqs.~(\ref{Eq:MM1}),(\ref{Eq:MM3}).

\section{NUMERICAL RESULTS AND DISCUSSIONS}\label{Sec6}

There are not any experimental data on the doubly charmed baryon
magnetic moments so far. We do not have any experimental inputs to
fit the LECs. In this paper, we use quark model to estimate the
leading-order low energy constants. At the leading order
$\mathcal{O}(p^{1})$, there are two unknown LECs $a_{1,2}$. The
charge matrix $Q_{H}$ is not traceless which is different from that
in the case of the light baryons. Notice that the $a_1$ parts are
proportional to the light quark charge within the doubly charmed
baryon. The $a_2$ parts are the same for the three doubly charmed
baryons and arise solely from the two charm quarks.

At the quark level, the flavor and spin wave function of the
$\Xi_{cc}^{++}$ reads:
\begin{eqnarray}
|\Xi_{cc}^{++};\uparrow\rangle&=&\frac{1}{3\sqrt{2}}[2c\uparrow
c\uparrow u\downarrow-c\uparrow c\downarrow u\uparrow-c\downarrow
c\uparrow u\uparrow +2c\uparrow u\downarrow c\uparrow-c\downarrow
u\uparrow c\uparrow\nonumber\\&&-c\downarrow u\downarrow
c\downarrow+2u\downarrow c\uparrow c\uparrow-u\downarrow c\downarrow
c\downarrow-u\uparrow c\downarrow c\uparrow], \label{xiwavefunc}
\end{eqnarray}
where the arrows denote the third-components of the spin. Replacing
the $u$ quark by the $d$ and $s$ quark, we get the wave functions of
the $\Xi_{cc}^{+}$ and $\Omega_{cc}^{+}$ respectively. The magnetic
moments of the doubly charmed baryons in the quark model are the
matrix elements of the following operator in Eq.~(\ref{xiwavefunc}),
\begin{equation}
\vec{\mu}=\sum_i\mu_i\vec{\sigma}^i, \label{magmomen}
\end{equation}
where $\mu_i$ is the magnetic moment of the quark.
\begin{equation}
\mu_i={e_i\over 2m_i},\quad i=u,d,s.
\end{equation}
We adopt the $m_u=m_d=336$ MeV, $m_s=540$ MeV, $m_c=1660$ MeV as the
constituent quark masses and give the results in the second column
in Table~\ref{various orders Magnetic moments}. The light quark
magnetic moments contributes to the LEC $a_{1}$, which is
proportional to the light quark charge. The heavy quark magnetic
moments contributes to the LEC $a_{2}$, which are the same for the
three doubly charmed baryons. The magnetic moments of the three
doubly charmed baryons are given in the second column in
Table~\ref{various orders Magnetic moments}.

Up to $\mathcal{O}(p^{2})$, we need include both the leading
tree-level magnetic moments and the $\mathcal{O}(p^{2})$ loop
corrections. At this order, there exists only one new LEC $\tilde{g}_A$. We
also use the quark model to estimate $\tilde{g}_A$. Considering the
$\pi^{0}$ coupling at the hadron level,
\begin{eqnarray}
\mathcal{L}_{\Xi_{cc}^{++}\Xi_{cc}^{++}\pi^{0}}=-\frac{1}{2F_{0}}\frac{\tilde{g}_{A}}{2}\bar{\Xi}_{cc}^{++}
\gamma^{\mu}\gamma_{5}\partial_{\mu}\pi^{0}\Xi_{cc}^{++}.
\end{eqnarray}
At the quark level, the $\pi^{0}$ quark interaction reads
\begin{eqnarray}
\mathcal{L}_{\rm quark}=\frac{1}{2}g_{0}\bar{\Psi}_q
\gamma^{\mu}\gamma_{5}\partial_{\mu}\pi^{0}\Psi_q.
\end{eqnarray}
With the help of the flavor wave functions of $\Xi_{cc}^{++}$, we
obtain the matrix elements at the hadron level
\begin{eqnarray}
\langle\Xi_{cc}^{++},s=\frac{1}{2}\mid
i\mathcal{L}_{\Xi_{cc}^{++}\Xi_{cc}^{++}\pi^{0}}\mid\Xi_{cc}^{++},s=\frac{1}{2};\pi^{0}\rangle
\sim-\frac{1}{2F_{0}}\frac{\tilde{g}_{A}}{2}q_3,
\end{eqnarray}
and at the quark level,
\begin{eqnarray}
\langle\Xi_{cc}^{++},s=\frac{1}{2}\mid i\mathcal{L}_{\rm
quark}\mid\Xi_{cc}^{++},s=\frac{1}{2};\pi^{0}\rangle
\sim-\frac{1}{6}g_0q_3.
\end{eqnarray}
After comparison with the axial charge of the nucleon,
 \begin{eqnarray}
\frac{-\frac{1}{2}\frac{\tilde{g}_{A}}{2}}{\frac{-1}{6}g_0}=\frac{\frac{1}{2}g_{A}}{\frac{5}{6}g_0},\label{eq:ga}
\end{eqnarray}
one obtains $\tilde{g}_A=\frac{2}{5}g_A=0.5$. Thus, we obtain the numerical
results of $\mathcal{O}(p^{2})$ chiral loop corrections in the third
column in Table~\ref{various orders Magnetic moments}. We list the
numerical results of $\mathcal{O}(p^{2})$ magnetic moments of the
three doubly charmed baryons in the fourth column in
Table~\ref{various orders Magnetic moments}. We also compare the
numerical results of the magnetic moments when the chiral expansions
are truncated at $\mathcal{O}(p^{1})$ and $\mathcal{O}(p^{2})$
respectively in Table~\ref{various orders Magnetic moments}.

Up to $\mathcal{O}(p^{3})$, there are six unknown LECs: $a_{1,2}$,
$g_{h1}$, $d_{1,2,3}$. Unfortunately, we are not able to present
numerical results since it is impossible to to fix all these LECs
with the available experimental information. We present our
analytical results in Eqs. (\ref{eq:mu2Loop}),(\ref{eq:mu3Loop}) and
Table~\ref{Magnetic moments}. Our analytical results may be useful
to the possible chiral extrapolation of the lattice simulations of
the doubly charmed baryon electromagnetic properties.

\begin{table}
  \centering
\begin{tabular}{c|ccccc}
\toprule[1pt]\toprule[1pt] Baryons & $\mathcal{O}(p^{1})$ tree &
$\mathcal{O}(p^{2})$ loop & $\mathcal{O}(p^{3})$ tree  &
$\mathcal{O}(p^{3})$ loop  \tabularnewline \midrule[1pt]
$\Xi_{cc}^{++}$ & $\frac{2}{3}a_{1}+4a_{2}$ & $-0.51\tilde{g}_{A}^2$ &
$-\frac{1}{3}d_{1}$ & $0.15a_{1}+0.21a_{2}-0.27g_{h1}$ &
\tabularnewline

$\Xi_{cc}^{+}$ & $-\frac{1}{3}$$a_{1}+4a_{2}$ &$0.15\tilde{g}_{A}^2$ &
$-\frac{1}{3}d_{1}$ & $-0.05a_{1}+0.21a_{2}+0.06g_{h1}$ &
\tabularnewline

$\Omega_{cc}^{+}$ & $-\frac{1}{3}a_{1}+4a_{2}$ & $0.36\tilde{g}_{A}^2$ &
$-\frac{1}{3}d_{1}-\frac{2}{3}d_{2}+4d_{3}$ &
$-0.12a_{1}+0.36a_{2}+0.21g_{h1}$ &  \tabularnewline
\bottomrule[1pt]\bottomrule[1pt]
\end{tabular}
\caption{The doubly charmed baryon magnetic moments to the
next-to-next-to-leading order(in unit of $\mu_{N}$).}
\label{Magnetic moments}
\end{table}

\begin{table}
  \centering
\begin{tabular}{c|cccc}
\hline\toprule[1pt]\toprule[1pt] Baryons& $\mathcal{O}(p^{1})$&
$\mathcal{O}(p^{2})$ loop & $\mathcal{O}(p^{2})$
total\tabularnewline \midrule[1pt] $\Xi_{cc}^{++}$ &
${4\over3}\mu_c-{1\over3}\mu_u=-0.12$ & $-0.13$ &
-0.25\tabularnewline \hline $\Xi_{cc}^{+}$ &
${4\over3}\mu_c-{1\over3}\mu_d=0.81$ & $0.04$ & 0.85\tabularnewline
\hline $\Omega_{cc}^{+}$ &  ${4\over3}\mu_c-{1\over3}\mu_s=0.69$ &
$0.09$ & 0.78\tabularnewline \bottomrule[1pt]\bottomrule[1pt]
\end{tabular}
\caption{The doubly charmed baryon magnetic moments when the chiral
expansion is truncated at $\mathcal{O}(p^{1})$ and
$\mathcal{O}(p^{2})$, respectively (in unit of $\mu_{N}$).}
\label{various orders Magnetic moments}
\end{table}

We also calculate the magnetic moments of the other doubly heavy baryons. At the quark level, the flavor and spin wave functions of the doubly bottomed baryons are the same as those of the doubly charmed baryons after replacing the c quarks by the b quarks. After the similar calculations of Eqs. (\ref{xiwavefunc})-(\ref{eq:ga}), one obtains the axial charge of doubly bottomed baryons $\tilde{g}_A(bbq)=\frac{2}{5}g_A$ and the tree level magnetic moments of the three doubly bottomed baryons in the second column in Table \ref{table:bbq}. We collect the numerical results of doubly bottomed baryon magnetic moments to next-to-leading order in Table \ref{table:bbq}.
\begin{table}
  \centering
\begin{tabular}{c|cccc}
\toprule[1pt]\toprule[1pt]
Baryons & $\mathcal{O}(p^{1})$ tree &  $\mathcal{O}(p^{2})$ loop & Total  \tabularnewline
\midrule[1pt]
$\Xi_{bb}^{0}$ & $\frac{4}{3}\mu_{b}-\frac{1}{3}\mu_{u}=-0.71$ & -0.13 & -0.84 \tabularnewline

$\Xi_{bb}^{-}$ & $\frac{4}{3}\mu_{b}-\frac{1}{3}\mu_{d}=0.22$ & 0.04 & 0.26 \tabularnewline

$\Omega_{bb}^{-}$ & $\frac{4}{3}\mu_{b}-\frac{1}{3}\mu_{s}=0.10$& 0.09 & 0.19  \tabularnewline
\bottomrule[1pt]\bottomrule[1pt]
\end{tabular}
\caption{The doubly bottomed baryon magnetic moments $(bbq)$ to the next-to-leading order(in
unit of $\mu_{N}$).} \label{table:bbq}
\end{table}

We also calculate the magnetic moments of the doubly heavy baryons containing a light quark, a charm quark and a bottom quark. We refer to the charm quark and the bottom quark as a diquark. There are two different multiplets of the doubly heavy baryons. The symmetric diquark $(\{bc\})$ has spin 1, while the antisymmetric diquark $([bc])$ has spin 0.

At the quark level, the flavor and spin wave function of the $(\{bc\}q)$ baryons reads,
\begin{eqnarray}
|\{bc\}q;\uparrow\rangle  =  \frac{1}{\sqrt{2}}(\mid cbq\rangle+\mid bcq\rangle)\otimes\frac{1}{\sqrt{6}}
(2\mid\uparrow\uparrow\downarrow\rangle-\mid\uparrow\downarrow\uparrow\rangle-\mid\downarrow\uparrow\uparrow\rangle),
\end{eqnarray}
while the flavor and spin wave function of the $([bc]q)$ baryons reads,
\begin{eqnarray}
|[bc]q;\uparrow\rangle  =  \frac{1}{\sqrt{2}}(\mid cbq\rangle-\mid bcq\rangle)\otimes\frac{1}{\sqrt{2}}
(\mid\uparrow\downarrow\uparrow\rangle-\mid\downarrow\uparrow\uparrow\rangle).
\end{eqnarray}
After the similar calculations, one obtains the axial charge of the $\{bc\}q$ baryons $\tilde{g}_A(\{bc\}q)=\frac{2}{5}g_A$ and the axial charge of the $[bc]q$ baryons $\tilde{g}_A([bc]q)=-\frac{6}{5}g_A$. We collect the tree level magnetic moments of the $\{bc\}q$ baryons in the second column in Table \ref{table:{bc}q} and the tree level magnetic moments of the three $[bc]q$ baryons in the second column in Table \ref{table:[bc]q}. We collect the numerical results of the $\{bc\}q$ and $[bc]q$ baryon magnetic moments to next-to-leading order in the fourth column in Table \ref{table:{bc}q} and Table \ref{table:[bc]q} respectively.

\begin{table}
  \centering
\begin{tabular}{c|cccc}
\toprule[1pt]\toprule[1pt]
Baryons & $\mathcal{O}(p^{1})$ tree &  $\mathcal{O}(p^{2})$ loop & Total  \tabularnewline
\midrule[1pt]
$\Xi_{\{bc\}q}^{+}$ & $\frac{1}{3}(2\mu_{b}+2\mu_c-\mu_u)=-0.41$ & -0.13 & -0.54 \tabularnewline

$\Xi_{\{bc\}q}^{0}$ & $\frac{1}{3}(2\mu_{b}+2\mu_c-\mu_d)=0.52$ & 0.04 & 0.56 \tabularnewline

$\Omega_{\{bc\}u}^{0}$ & $\frac{1}{3}(2\mu_{b}+2\mu_c-\mu_s)=0.40$& 0.09 & 0.49  \tabularnewline
\bottomrule[1pt]\bottomrule[1pt]
\end{tabular}
\caption{The magnetic moments of doubly heavy baryons $(\{bc\}q)$ to the next-to-leading order(in
unit of $\mu_{N}$).} \label{table:{bc}q}
\end{table}

\begin{table}
  \centering
\begin{tabular}{c|cccc}
\toprule[1pt]\toprule[1pt]
Baryons& $\mathcal{O}(p^{1})$ tree &  $\mathcal{O}(p^{2})$ loop & Total  \tabularnewline
\midrule[1pt]
$\Xi_{[bc]q}^{+}$ & $\mu_u=1.86$ & -1.17 & 0.69 \tabularnewline

$\Xi_{[bc]q}^{0}$ & $\mu_d=-0.93$ & 0.34 & -0.59 \tabularnewline

$\Omega_{[bc]u}^{0}$ & $\mu_s=-0.58$& 0.82 & 0.24  \tabularnewline
\bottomrule[1pt]\bottomrule[1pt]
\end{tabular}
\caption{The magnetic moments of doubly heavy baryons $([bc]q)$ to the next-to-leading order(in
unit of $\mu_{N}$).} \label{table:[bc]q}
\end{table}

\section{Conclusions}\label{Sec7}

The discovery of the $\Xi^{++}_{cc}$ inspired heated theoretical
investigation of the doubly charmed baryons. The doubly charmed
baryons are so special that the chiral dynamics is dominated by the
single light quark. The electromagnetic property of the doubly
charmed baryons encodes crucial information of their inner
structure. In this work, we have performed a systematical
calculations of the chiral corrections to the magnetic moments of
doubly charmed baryons up to the next-to-next-to-leading order in
the framework of heavy baryon chiral perturbation theory. We use
quark model to estimate the low energy constants and present the
numerical results up to next-to-leading order:
$\mu_{\Xi^{++}_{cc}}=-0.25\mu_{N}$,
$\mu_{\Xi^{+}_{cc}}=0.85\mu_{N}$,
$\mu_{\Omega^{+}_{cc}}=0.78\mu_{N}$.

From Table~\ref{various orders Magnetic moments}, the magnetic
moments of the $\Xi^{+}_{cc}$ and $\Omega^{+}_{cc}$ are dominated by
the leading order term while the chiral corrections are quite small.
To be specific, the numerical values of the $\mathcal{O}(p^{1})$
magnetic moments of the $\Xi^{+}_{cc}$ and $\Omega^{+}_{cc}$ are
enhanced since the charge of the down and strange quark is $-{1\over
3}$ while the charm quark charge is $+{2\over 3}$. Only the $\pi^+$
meson contributes to the chiral correction to $\mu_{\Xi^{+}_{cc}}$
at $\mathcal{O}(p^{2})$ while only $K^+$ contributes to
$\mu_{\Omega^{+}_{cc}}$ at this order.

For comparison, the up and charm quark contributions to the
$\mathcal{O}(p^{1})$ magnetic moment of the $\Xi^{++}_{cc}$ are
destructive. Such an accidental strong cancelation renders the
leading order magnetic moment of the $\Xi^{++}_{cc}$ is much smaller
than those of its partner states. In contrast, both the $\pi^+$ and
$K^+$ mesons contribute to the chiral corrections to
$\mu_{\Xi^{++}_{cc}}$ at $\mathcal{O}(p^{2})$. In other words, the
leading order magnetic moment of the $\Xi^{++}_{cc}$ is reduced
while the loop correction is enhanced. As a result, the loop
correction is numerically very important and even slightly larger
than the leading order term. Such a unique feature can be exposed by
future lattice QCD simulation.

In Table~\ref{Comparison of magnetic moments}, we compare our
results obtained in the HBChPT with those from other model
calculations such as quark model (QM)~\cite{Lichtenberg:1976fi},
relativistic three-quark model (RTQM)~\cite{Faessler:2006ft},
nonrelativistic quark model in Faddeev approach
(NQM)~\cite{SilvestreBrac:1996bg}, relativistic quark model
(RQM)~\cite{JuliaDiaz:2004vh}, skyrmion
description~\cite{Oh:1991ws}, confining logarithmic potential
(CLP)~\cite{Jena:1986xs}, MIT bag model~\cite{Bose:1980vy},
nonrelativistic quark model (NQM)~\cite{Patel:2008xs} and lattice
QCD(LQCD). All these approaches lead to roughly consistent results.

As the byproducts, we have also calculated the magnetic moments of
the other doubly heavy baryons, including the $bbq$ baryons, the
$\{bc\}q$ baryons and the $[bc]q$ baryons. Especially, the magnetic
moments of $[bc]q$ baryons are quite interesting as their magnetic
moments totally arise from the light quarks as shown in Table
\ref{table:[bc]q}.

We hope our calculation may be useful for future experimental
measurements. As there are several unknown LECs up to
next-to-next-to-leading order, we are looking forward to further
progresses in both theory and experiment so that we can check the
chiral expansion convergence of the three doubly charmed baryons.
Our results may be useful for future experimental measurement of the
magnetic moments. Our analytical results may also be useful to the
possible chiral extrapolation of the lattice simulations.

 \begin{table}
  \centering
\begin{tabular}{c|ccc}
\toprule[1pt]\toprule[1pt] Baryons & $\Xi_{cc}^{++}$ &
$\Xi_{cc}^{+}$ & $\Omega_{cc}^{+}$\tabularnewline \midrule[1pt]
QM~\cite{Lichtenberg:1976fi} & -0.124 & 0.806 & 0.688\tabularnewline
\hline RTQM \cite{Faessler:2006ft} & 0.13 & 0.72 &
0.67\tabularnewline \hline NRQM \cite{SilvestreBrac:1996bg} & -0.206
& 0.784 & 0.635\tabularnewline \hline RQM \cite{JuliaDiaz:2004vh} &
-0.10 & 0.86 & 0.72\tabularnewline \hline Skyrmion \cite{Oh:1991ws}
& -0.47 & 0.98 & 0.59\tabularnewline \hline CLP \cite{Jena:1986xs} &
-0.154 & 0.778 & 0.657\tabularnewline \hline MIT bag model
\cite{Bose:1980vy} & 0.17 & 0.86 & 0.84\tabularnewline \hline NQM
\cite{Patel:2008xs} & -0.208 & 0.785 & 0.635\tabularnewline \hline
LQCD \cite{Can:2013tna} & \textemdash{} & 0.425 &
0.413\tabularnewline \hline This work & -0.25 & 0.85 &
0.78\tabularnewline \bottomrule[1pt]\bottomrule[1pt]
\end{tabular}
\caption{Comparison of the decuplet to octet baryon transition
magnetic moments in literature including quark model
(QM)~\cite{Lichtenberg:1976fi}, relativistic three-quark model
(RTQM)~\cite{Faessler:2006ft}, nonrelativistic quark model in
Faddeev approach (NQM)~\cite{SilvestreBrac:1996bg}, relativistic
quark model (RQM)~\cite{JuliaDiaz:2004vh}, skyrmion
description~\cite{Oh:1991ws}, confining logarithmic potential
(CLP)~\cite{Jena:1986xs}, MIT bag model~\cite{Bose:1980vy},
nonrelativistic quark model (NQM)~\cite{Patel:2008xs} and lattice
QCD(LQCD) \cite{Can:2013tna}(in unit of $\mu_{N}$).}
  \label{Comparison of magnetic moments}
 \end{table}

\section*{ACKNOWLEDGMENTS}

H. S. Li is very grateful to X. L. Chen and W. Z. Deng for very
helpful discussions. This project is supported by the National
Natural Science Foundation of China under Grants 11575008,
11621131001 and 973 program. This work is also supported by the
Fundamental Research Funds for the Central Universities of Lanzhou
University under Grants 223000--862637.

\begin{appendix}

%
%

\section{COEFFICIENTS OF THE LOOP CORRECTIONS} \label{appendix-B}

In this appendix, we collect the explicit formulae for the chiral
expansion of the doubly charmed baryon magnetic moments in Tables
\ref{table:abd} and \ref{table:ef}.

\begin{table}
  \centering
\begin{tabular}{c|ccccccc}
\toprule[1pt]\toprule[1pt] Baryons & $\beta_{a}^{\pi}$ &
$\beta_{a}^{K}$ & $\beta_{b}^{\pi}$ & $\beta_{b}^{K}$ &
$\beta_{d}^{\pi}$ & $\beta_{d}^{K}$ &
$\beta_{d}^{\eta}$\tabularnewline \midrule[1pt] $\Xi_{cc}^{++}$ &
$2$ & $2$ & $-4a_{1}$ & $-4a_{1}$ & $24a_{2}$ &
$-\frac{4}{3}a_{1}+16a_{2}$ &
$\frac{4}{9}(a_{1}+6a_{2})$\tabularnewline \hline $\Xi_{cc}^{+}$ &
$-2$ & 0 & $4a_{1}$ & 0 & $2a_{1}+24a_{2}$ &
$-\frac{4}{3}a_{1}+16a_{2}$ &
$\frac{2}{9}(-a_{1}+12a_{2})$\tabularnewline \hline
$\Omega_{cc}^{+}$ & 0 & $-2$ & 0 & $4a_{1}$ & 0 &
$\frac{4}{3}(a_{1}+24a_{2})$ &
$-\frac{8}{9}(a_{1}-12a_{2})$\tabularnewline
\bottomrule[1pt]\bottomrule[1pt]
\end{tabular}
\caption{The coefficients of the loop corrections to the
 doubly charmed baryon magnetic moments from Figs.
\ref{fig:allloop}(a), \ref{fig:allloop}(b) and
\ref{fig:allloop}(d).} \label{table:abd}
\end{table}

\begin{table}
  \centering
\begin{tabular}{c|c|c|c|c|c}
\toprule[1pt]\toprule[1pt] Baryons & $\beta_{e}^{\pi}$ &
$\beta_{e}^{K}$ & $\beta_{f}^{\pi}$ & $\beta_{f}^{K}$ &
$\beta_{f}^{\eta}$\tabularnewline \midrule[1pt] $\Xi_{cc}^{++}$ &
$g_{h1}$ & $g_{h1}$ & $2(a_{1}+6a_{2})$ &
$\frac{4}{3}(a_{1}+6a_{2})$ &
$\frac{2}{9}(a_{1}+6a_{2})$\tabularnewline \hline $\Xi_{cc}^{+}$ &
$-g_{h1}$ & 0 & $-a_{1}+12a_{2}$ & $-\frac{2}{3}a_{1}+8a_{2}$ &
$-\frac{1}{9}a_{1}+\frac{4}{3}a_{2}$\tabularnewline \hline
$\Omega_{cc}^{+}$ & 0 & $-g_{h1}$ & 0 & $-\frac{4}{3}a_{1}+16a_{2}$
& $-\frac{4}{9}(a_{1}-12a_{2})$\tabularnewline
\bottomrule[1pt]\bottomrule[1pt]
\end{tabular}
\caption{The coefficients of the loop corrections to the
 doubly charmed baryon magnetic moments from Figs.
\ref{fig:allloop}(e) and \ref{fig:allloop}(f).} \label{table:ef}
\end{table}

\end{appendix}

\vfil \thispagestyle{empty}

\end{document}